\documentclass[%
 reprint,
 amsmath,amssymb,
 aps,
]{revtex4}

\usepackage{graphicx}% Include figure files
\usepackage{dcolumn}% Align table columns on decimal point
\usepackage{bm}% bold math
\begin{document}

\preprint{APS/123-QED}

\title{Super Penrose Process for Kerr Black Hole And White Hole in Rainbow Gravity}% Force line breaks with \\
%\thanks{A footnote to the article title}%

\author{Shun Jiang}

\email{shunjiang@mail.bnu.edu.cn}

\affiliation{%
	Department of Physics, Beijing Normal University, Beijing, 100875, China\\}%Lines break automatically or can be forced with \\

\date{\today}% It is always \today, today,
             %  but any date may be explicitly specified

\begin{abstract}
Recently, O.B.Zaslavskii considers the collision near the horizon of the extremal Kerr black hole that produces two particles and one of two particles has divergent conserved killing energy(the so called super Penrose process). O.B.Zaslavskii shows this particle can not move in a asymptotic region in this universe. This means super Penrose process for extremal Kerr black hole is impossible. However, recently, Mandar Patil and Tomohiro Harada show this particle can enter event horizon, turns back and emerges through white hole event horizon into asymptotic region in other universe. This means super Penrose process for extremal Kerr white hole is possible. During this progress, they assume this high energy particle's trajectory is geodesic in Kerr background. However, according to rainbow gravity, an unbounded energy particle will change the spacetime background and observe different effective space-time geometries. Furthermore, this particle's angular momentum is slightly less than the max value which particle can enter black hole. Therefore, when one considers rainbow gravity, this particle with unbounded energy may not enter black hole. In this paper, we will consider rainbow gravity's effect. We will show for some rainbow functions, there is a critical energy. If particle's energy is greater than it, particle will not enter black hole and turn back. It means super Penrose process for extremal Kerr white hole becomes impossible in some rainbow gravity. Because of this high energy particle turning back, the super Penrose process for extremal Kerr black hole becomes possible. Therefore, we see the possibility of super Penrose process for Kerr black hole or white hole depends on rainbow gravity. 
\end{abstract}

%\keywords{Suggested keywords}%Use showkeys class option if keyword
                              %display desired
\maketitle

%\tableofcontents

\section{introduction}
In last decade, high energy processes near black hole have aroused wide attention. In \cite{Ba_ados_2009}, they show energy in the center of mass $E_{c.m.}$ can grow unbounded if one of particles has a critical value of angular momentum. This was called BSW effect. After it publication, a lot of high energy collisions has been studied \cite{zaslavskii2020special,hackmann2020particle,Jiang_2019,Gonz_lez_2018,an2017naked,An_2018,Amir_2016,wang2015bsw,Li_2011,Zhang:2018gpn,Zhang:2020cpu,Guo:2016vbt}. They focus on energy in the center of mass. An interesting question is whether it is possible to gain an unbounded killing energy $E$. This is called the super Penrose process. If a particle with an unbounded killing energy can be produced after collsions, the Earth laboratory can measure it in principle. An unbounded killing energy means particle need extract energy from black hole during collisions. For such an extraction, the negative energy orbits and ergoregion play a key role. It is obvious the Kerr black hole satisfies these conditions. However, in \cite{Zaslavskii_2016}, Zaslavskii O. B. show the unbounded killing energy particle produced by collision can not move in a asymptotic region in this universe and they all fall into black hole. So this progress is impossible for kerr black hole.

Recently, another scenarion was considered for the extremal Kerr black hole\cite{Patil_2020}. They consider the collision happens near black hole and produces two massless particles. They need one massless particle with unbounded killing energy enters black hole, turns back and appears in a white hole region. As a result, high energy fluxes can observed in another universe. Or, vice versa, if the collision happens in some other universe, this leads to high particles in our universe and can be observed in principle. Therefore, super Penrose process for kerr white hole is possible. An important step for this progress is that the massless particle need enter black hole. This depends on particle's energy and angular momentum. For extremal Kerr black hole, this become $L<2ME$ where $L$, $E$ and $M$ are partcile's angular momentum, energy and black hole mass. However, as we will see, for massless particle produced in this progress, the angular momentum $L$ is very close to $2ME$. This means if there is a little change on the background, this particle may not enter black hole. In fact, in this situation, we consider a particle with unbounded killing energy. According to rainbow gravity, different energy test particles observe different effective space-time geometries. This unbounded killing energy gives a considerable effect on particle's trajectory in rainbow gravity. Therefore, it is nature to consider the rainbow gravity's effect on particle's trajectory. This effect may prevent particle entering black hole and changes super Penrose process for Kerr white hole.

Rainbow gravity is a generalization of modified dispersion relations in curved spacetimes. It was proposed by Magueijo and Smolin in 2004 \cite{Magueijo_2004}. In this theory, the geometry of spacetime depends on the energy of the test particle. This means two different test particles observe different effective space-time geometries. It goes back to classical general relativity in the low energy limit. In this collision progress, we are interested in the massless particle with unbounded energy. It is reasonable to believe the rainbow gravity will give some considerable effects. So in this paper, we will consider this massless particle's trajectory in rainbow gravity. We will show there are some rainbow functions prevent this massless particle entering black hole when energy is higher than a critical value. This means super Penrose process for kerr white hole becomes impossible in rainbow gravity. However, at the same time, this massless particle will turn back and move in an asymptotic region in our universe. Therefore, rainbow gravity makes super Penrose process for kerr black hole become possible. We see the rainbow gravity changes the possibility of super Penrose process for Kerr black hole and white hole. 

There are several reasons for considering rainbow gravity. First, we consider a high energy progress where the collision produces an unbounded killing energy massless partcile. If particle's energy is very high, the rainbow gravity effects will become strong and can not be ignored. Second, as we will see, this particle's angular momentum is slightly less than the max angular momentum which particle can enter black hole. Therefore, a little change on background may prevent this massless particle enter black hole. This change can be discribed by rainbow gravity. Third, it will be a test on rainbow gravity in principle. As we will see, for some rainbow functions, there is an upper bounded energy for massless particle, if the massless particle's energy greater than it, the massless particle can not enter event horizon. This particle will turns back and move in an asymptotic region in our universe. If one observe massless particle's energy greater than it in this progress, this may help us to exclude some rainbow functions. Or, vice versa, if the energy observed all less than upper bounded energy, this may be an evidence for rainbow gravity.

This paper is organized as follows: In Sec.~\ref{section1} we discuss the super Penrose process in extremal Kerr white hole. In Sec.~\ref{s2}, we discuss rainbow gravity's effect on super Penrose proceess. In Sec.~\ref{section5}, we briefly summarize our results.
\section{super penrose progress in extremal kerr white hole}\label{section1}
In this section, we review the super Penrose process for Kerr white hole. Firstly, let us show the physical picture of this progress: we consider the collision of two particles 1 and 2 near the horizon of extremal Kerr black hole and prducing particles 3 and 4. We need both particles enter black hole, turn back and move in a white hole region. If particle 1 has critical angular momentum, the particle 3 has an unbounded Killing energy $E$ and it can reach a flat infinity in another universe. This can be detected by laboratory in principle and white hole can be regarded as the source of ultra high energy particles. In \cite{Patil_2020}, the authors consider this picture by using transformation between three frames (stationary one, locally non-rotating and center of mass). We prefer to consider this problem in the original frame. We work in Boyer Lindquist coordinate system. An important aim of section is getting particle 3's energy and angular momentum. We will see this angular momentum is slightly less than the max angular momentum which particle can enter black hole. 

Extremal Kerr black hole metric can be written as
\begin{eqnarray}
	ds^2=-(1-\frac{2Mr}{\Sigma})dt^2+\frac{\Sigma}{\Delta}dr^2+\Sigma d\theta^2  \nonumber \\
	-\frac{4M^2r}{\Sigma}\sin^2\theta d\varphi dt+(r^2+M^2+\frac{2Mr}{\Sigma}\sin^2\theta)d\varphi^2
\end{eqnarray}
where $\Sigma$ and $\Delta$ satisfy
\begin{eqnarray}
	\Sigma=r^2+M^2\cos^2\theta \nonumber \\
	\Delta=(r-M)^2
\end{eqnarray}
We use $U^a$ to present massive particle's four volocity. For a massive particle, the geodesic equation on the equatorial plane can be wirtten as
\begin{eqnarray}
	\frac{dt}{d\tau}=\frac{1}{(r-M)^2}[E(r^2+M^2+\frac{2M^3}{r})-\frac{2M^2L}{r}]
\end{eqnarray}
\begin{eqnarray}
			\frac{d\varphi}{d\tau}=\frac{1}{(r-M)^2}[\frac{2M^2E}{r}+(1-\frac{2M}{r})L]
\end{eqnarray}
$E$ and $L$ are the killing energy and angular momentum. They can be expressed as
\begin{eqnarray}
E=-U_a(\frac{\partial}{\partial t})^a\\
L=U_a(\frac{\partial}{\partial \varphi})^a
\end{eqnarray}
We are interested in the situation where particle falls inward starting from rest at infinity. Therefore, we set $E=1$. The radial part motion can be written as
\begin{eqnarray}
	\frac{dr}{d\tau}=\sigma\sqrt{\frac{2M}{r}-\frac{L^2}{r^2}+\frac{2M(L-M)^2}{r^3}} \label{r}
\end{eqnarray}
For a massless particle, we consider geodesic motion on the equatorial plane.The massless particle's four volocity $k^a$ can be wirtten as
\begin{eqnarray}
	k^t=\frac{1}{(r-M)^2}[E(r^2+M^2+\frac{2M^3}{r})-\frac{2M^2L}{r}]
\end{eqnarray}
\begin{eqnarray}
k^{\varphi}=\frac{1}{(r-M)^2}[\frac{2ME}{r}+(1-\frac{2M}{r})L]
\end{eqnarray}
\begin{eqnarray}
	k^r=\sigma\sqrt{E^2-\frac{(L^2-M^2E^2)}{r^2}+\frac{2M(L-ME)^2}{r^3}}
\end{eqnarray}
where $E$ and $L$ are the killing energy and angular momentum.

We focus on the massless particle which can enter black hole and move in a white hole region. This condition can be find in \cite{Patil_2020}. In order to enter the black hole, we need $k^t>0$ at horizon, which imply that $2ME>L$. In order to appear in a white hole region, we need a turning point inside black hole, which imply $ME<L$. Therefore, if $ME<L<2ME$, the massless particle will enter the black hole and emerge out of white hole region.

Now, we consider two identical particles with unit mass($m=1$) which fall towards the black hole starting from rest at infinity. Therefore, for these particles, the killing energy are $E_1=1$ and $E_2=1$. We assume the particle 1 has critical angular momentum $L_1=2M$ so that particle 1 will asymptotically approaches the event horizon loacted at $r=M$. While the particle 2 has the angular momentum in the range $-2(1+\sqrt{2})M<L<2M$ so that this particle can arrive the horizon with finite radial component of velocity \cite{Patil_2020}. We asuume both particles travel in the inward direction so that $\sigma_1=\sigma_2=-1$ in Eqs.~(\ref{r}). We consider the collision occurs at $r=M(1+\epsilon)$ which is very close to the event horizon and produing massless particle 3 and 4. During this progress, four momentum are conserved. We have $P^a_1+P^a_2=P^a_3+P^a_4$ at $r=M(1+\epsilon)$. Where $P^a_i$ means particle $i$'s four momentum.

For massive particle with unit mass($m=1$), $P^t$ and $P^r$ at $r=M(1+\epsilon)$ can be written as
\begin{eqnarray}
	P^t=\frac{1}{M^2\epsilon^2}[4M^2+3M^2\epsilon^2-2M(1-\epsilon+\epsilon^2)L]\\
	P^r=\sigma\sqrt{(2-\bar{L})^2+4(2-\bar{L})(\bar{L}-1)\epsilon}
\end{eqnarray}
where we define $\bar{L}=L/M$. For particle 1 with killing energy $E=1$ and critical angular momentum $L_1=2M$, we have
\begin{eqnarray}
	P^t_1=\frac{4}{\epsilon}+1\\
	P^r_1=-\sqrt{2}\epsilon
\end{eqnarray}
For particle 2 with killing energy $E_2$ and angular momentum $L_2$, we have
\begin{eqnarray}
	P^t_2=\frac{4-2\bar{L}_2}{\epsilon^2}+\frac{2\bar{L}_2}{\epsilon}\\
    P^r_2=-(2-\bar{L}_2)-2(\bar{L}_2-1)\epsilon
\end{eqnarray}
For massless particle, $P^t$ and $P^r$ at $r=M(1+\epsilon)$ can be written as
\begin{eqnarray}
	P^t=\frac{1}{(r-M)^2}[(r^2+M^2+\frac{2M^3}{r})E-\frac{2M^2}{r}L]\\
    P^r=-\sqrt{A^2+B\epsilon+C\epsilon^2+O(\epsilon^3)} \label{radial}
\end{eqnarray}
where $A=(2E-\bar{L})$, $B=-4(\bar{L}-E)(\bar{L}-2E)$, $C=3(\bar{L}-E)(3\bar{L}-5E)$.
we need four momentum conserve $P^a_1+P^a_2=P^a_3+P^a_4$. It becomes
\begin{eqnarray}
	E_1+E_2=2=E_3+E_4\\
	\bar{L}_1+\bar{L}_2=\bar{L}_3+\bar{L}_4\\
		P^r_1+P^r_2=-2+\bar{L}_2-(2\bar{L}_2-2+\sqrt{2})\epsilon=P^r_3+P^r_4 \label{a1}
\end{eqnarray}
Let us focus on the photon's radial momentum
\begin{eqnarray}
	P^r=-\sqrt{A^2+B\epsilon+C\epsilon^2+O(\epsilon^3)} 
\end{eqnarray}
In order to have unbound energy, we need photon $E_3\sim 1/\sqrt{\epsilon}$. According to Eqs.~(\ref{radial}) and Eqs.~(\ref{a1}), we need $2E_3-\bar{L}_3$ is finite. Because the left hand of Eqs.~(\ref{a1}) is finite. Therefore, we can assume $\bar{L}_3=2N_3/\sqrt{\epsilon}$, $E_3=X_3+N_3/\sqrt{\epsilon}$, where $N_3$ and $X_3$ are constants. In order to enter the black hole and move in a white hole region, we need $P^t_3>0$ at horizon and this leads to $X_3>0$. For the same reason, the killing energy and angular momentum of particle 4 can be written as $E_4=X_4+N_4/\sqrt{\epsilon}$ and $\bar{L}_4=2N_4/\sqrt{\epsilon}$. We are interested in both massless particle enter black hole. Therefore, we need $X_4>0$. Using the explicit form of $E$ and $L$, the photon's radial momentum can be written as
\begin{eqnarray}
	P^r=2X\sqrt{1+\frac{2N}{X}\sqrt{\epsilon}+\frac{3N^2-8X^2}{4X^2}\epsilon} \nonumber\\ \label{Pr}
\end{eqnarray}
Defining $x=\sqrt{\epsilon}$ and using taylor expansion to second order, we have
\begin{eqnarray}
	\sqrt{1+ax+bx^2}=1+\frac{a}{2}x+(\frac{b}{2}-\frac{a^2}{8})x^2
\end{eqnarray}
The photon's radial momentum becomes
\begin{eqnarray}
	P^r=2X+2N\sqrt{\epsilon}-(\frac{N^2}{4X}+2X)\epsilon
\end{eqnarray}
Energy and angular momentum conserve becomes
\begin{eqnarray}
	X_3+X_4+\frac{N_3}{\sqrt{\epsilon}}+\frac{N_4}{\sqrt{\epsilon}}=2 \label{E}\\
	\frac{2N_3}{\sqrt{\epsilon}}+\frac{2N_4}{\sqrt{\epsilon}}=2+\bar{L}_2 \label{L}
\end{eqnarray}
Using Eqs.~(\ref{E}) and Eqs.~(\ref{L}), Eqs.~(\ref{a1}) can be written as
\begin{eqnarray}
	\frac{N^2_3}{4X_3}+\frac{N^2_4}{4X_4}=2-\sqrt{2}
\end{eqnarray}
Using $N_3\sim -N_4$, Eqs.~(\ref{E}) and Eqs.~(\ref{L}), We obtain the expression of $X_{3,4}$
\begin{eqnarray}
X_{3,4}=\frac{AB\mp\sqrt{AB}\sqrt{AB-4}}{2B}\label{34}
\end{eqnarray}
where $A=1-\bar{L}_2/2$ and $B=(8-4\sqrt{2})/N^2_3$. We need $AB>4$. It is easy to see $X_{3,4}>0$ in this situation. This means two massless particles can enter balck hole and particle 3 with positive unbounded killing energy can turn back and move in a white hole region.

\section{super penrose progress in extremal kerr white hole in rainbow gravity}\label{s2}
In previous section,  we consider the collision of two particles 1 and 2 near the horizon of extremal kerr black hole and producing particles 3 and 4. We need both particles enter black hole and move in a white hole region. We show particle 3 with unbounded killing energy can enter black hole and move in a white hole region. 

However, there are some problems. First, in this picture, we are interested in the unbounded high killing energy particle. It is doubtful whether a particle with such high energy travels the geodesic trajectory in kerr background. In fact, from Einstein field equation, this high energy may change the backgroud and change particle's trajectory. Second, we see the particle's angular momentum and energy can be expressed as $\bar{L}_3=2N_3/\sqrt{\epsilon}$, $E_3=X_3+N_3/\sqrt{\epsilon}$. The condition for particle to enter black hole is $2E>\bar{L}$. However, when $\epsilon \rightarrow 0$, the angular momentum $\bar{L}$ is very close to $2E$. Therefore, if high enery change the condition for particle to enter black hole slightly, it may prevent this high energy particle entering black hole. 

Rainbow gravity gives us a tool to take into account the particle's energy effect on particle's trajectory. In this situation, we are interested in an unbounded high energy particle. Therefore, we think rainbow gravity will give a considerable effect on particle's trajactory. Because of these reasons, in the rest of this section, we will check whether particle with such high energy can enter black hole in rainbow gravity. 

The Kerr metric in rainbow gravity reads \cite{Dubey_2017}
\begin{eqnarray}
	ds^2&&=\frac{g_{tt}}{f^2(E/E_{pl})}dt^2+\frac{g_{\varphi\varphi}}{g^2(E/E_{pl})}d\varphi^2+\frac{g_{rr}}{g^2(E/E_{pl})}dr^2\nonumber \\
	&&+\frac{2g_{t\varphi}}{f(E/E_{pl})g(E/E_{pl})}dtd\varphi +\frac{g_{\theta\theta}}{g^2(E/E_{pl})}d\theta^2
\end{eqnarray}
where $g_{\mu\nu}$ are Kerr metric components in Boyer Lindquist coordinate system. There are two rainbow function $f(E/E_{pl})$ and $g(E/E_{pl})$ in this metric. This means particle with different energy $E$ will observe different effective spacetime geometries. In the low energy limit, we have
\begin{eqnarray}
	\lim\limits_{E/E_{pl}\rightarrow \infty}f(E/E_{pl})=\lim\limits_{E/E_{pl}\rightarrow \infty}g(E/E_{pl})=1
\end{eqnarray}
The metric will go back to the Kerr metric in the low energy limit. In this time, we are consider a particle with unbounded high killing energy. We believe rainbow gravity will give some considerable effects.

First, we need calculate the condition for massless particle entering black hole in this metric. The $t$ component of massless particle's four volocity $k^a$ can be written as
\begin{eqnarray} 
	k^t=\frac{Ef^2(r^2+M^2+\frac{2M^3}{r})-\frac{2M^2Lfg}{r}}{(r-M)^2}
\end{eqnarray}
If the massless particle can enter black hole , we need $k^t>0$ at $r=M$. This leads to
\begin{eqnarray}
	\bar{L}<2E\frac{f(E/E_{pl})}{g(E/E_{pl})} \label{enter}
\end{eqnarray}
Let us focus on this condition. It is easy to see this condition goes back to the Kerr one in the low energy limit. There are three 3 cases for rainbow functions. First, when $f(E/E_{pl})=g(E/E_{pl})$, the condition Eqs.~(\ref{enter}) goes back to Kerr case. Therefore,  the massless particle can enter black hole. Second, when $f(E/E_{pl})>g(E/E_{pl})$, the massless particle also can enter black hole. Third, when $f(E/E_{pl})<g(E/E_{pl})$, the massless particle may not enter black hole in rainbow gravity. 

Let us consider a explicit example for the third case. From \cite{LEIVA_2009}, we read rainbow function: $f(E)=1$, $g(E)=1+1/2l_pE$, where $l_p$ is plank length.
Taking them into Eqs.~(\ref{enter}) and using the explicit form of particle 3's $E$ and $\bar{L}$, we find
\begin{eqnarray}
	2X_3+\frac{2N_3}{\sqrt{\epsilon}}>\frac{2N_3}{\sqrt{\epsilon}}[1+\frac{1}{2}l_p(X_3+\frac{N_3}{\sqrt{\epsilon}})]
\end{eqnarray}
let $y=\sqrt{\epsilon}$. It becomes 
\begin{eqnarray}
	2X_3y^2-N_3l_pX_3y-N_3^2l_p>0
\end{eqnarray}
we find
\begin{eqnarray}
	y>\frac{N_3l_pX_3+\sqrt{N_3^2l^2_pX_3^2+8N_3^2l_pX_3}}{4X_3}
\end{eqnarray}
we find an upper bound for energy
\begin{eqnarray}
	E<E_c=X_3+\frac{4N_3X_3}{N_3l_pX_3+\sqrt{N_3^2l^2_pX_3^2+8N_3^2l_pX_3}}
\end{eqnarray}
where we use $E_c$ to represent the upper boundary energy.

If massless particle's energy greater than $E_c$, then this particle can not enter black hole. This means super Penrose progress for Kerr white hole becomes impossible in rainbow gravity. At the same time, this massless particle with energy greater than $E_c$ will turn back and move in a asymptotic region in this universe. So, super Penrose progress for Kerr black hole becomes possible in rainbow gravity. We see the possibility of super Penrose process for kerr black hole or white hole depends on rainbow gravity.

This result can also be used to test rainbow gravity. O.B.Zaslavskii shows super Penrose process for extremal Kerr black hole is impossible\cite{Zaslavskii_2016}. However, with rainbow gravity, this becomes possible for some rainbow functions($f(E/E_{pl})<g(E/E_{pl})$). If we observe super Penrose process for extremal Kerr black hole in laboratory, it will become an evidence for rainbow gravity and help us choose rainbow functions. Therefore, this result gives a method to test rainbow gravity in principle.

\section{conclusion}\label{section5}
In this paper, we study the super Penrose progress for Kerr white hole in rainbow gravity. We consider the rainbow effect on particle's trajectory. In\cite{Patil_2020}, they consider two identical massive particles fall inwards starting from rest at infinity towards the extremal Kerr black hole. they assume particles collide outside the event horizon in its vicinity and produce two massless particle. they show one massless particle will have unbounded killing energy and this massless particle wiil enter black hole, turn back and move in a white hole region. Therefore, Kerr white hole can be a potential source of ultra high energy particles and in principle, this high energy particles can be observed. This means super Penrose progress for Kerr white hole is possible. However, we find this unbounded energy particle has a sensitive angular momentum: its angular momentum is slightly less than the max angular momentum which particle can enter black hole. What's more, this particle has an unbounded energy. This unbounded energy may change background and prevents particle entering black hole. This change can be described by rainbow gravity. So we conisder the rainbow effect on particle's trajectory. We show some rainbow function can prevent particle entering black hole when particle's energy is greater than a critical value. Therefore, with this kind rainbow functions, super Penrose progress for Kerr white hole is impossible. At this times, this particle will turn back and move in a asymptotic region in this universe. This means super penrose progress for Kerr balck hole becomes possible in rainbow gravity which is impossible without rainbow gravity \cite{Zaslavskii_2016}. Therefore, we see the possibility of super Penrose process for Kerr black hole and white hole depends rainbow gravity. 
\begin{acknowledgments}
This research was supported by NSFC Grants No. 11775022 and 11873044.
\end{acknowledgments}

\appendix

\bibliography{spp}% Produces the bibliography via BibTeX.

\end{document}